\documentclass[aps,prd,twocolumn,superscriptaddress,nofootinbib,floatfix]{revtex4-2}

\usepackage{amsmath,amssymb}
\usepackage{graphicx}
\usepackage{hyperref}
\usepackage{xcolor}
\usepackage{bm}
\usepackage{multirow}
\usepackage{booktabs}
\usepackage{placeins}
\usepackage{orcidlink}
\usepackage{float}

\newcommand{\dLGW}{d_L^{\rm GW}}
\newcommand{\dLEM}{d_L^{\rm EM}}
\newcommand{\Om}{\Omega_m}
\newcommand{\ellz}{\ell_0}
\newcommand{\cT}{c_T}
\newcommand{\GT}{G_T}

\begin{document}

\title{Gravitational Wave Standard Sirens as Probes of Lorentz Violation
       in Bumblebee Gravity}

\author{Ayan Banerjee \orcidlink{0000-0003-3422-8233}}
\email{ayanbanerjeemath@gmail.com}
\affiliation{Astrophysics and Cosmology Research Unit, School of Mathematics, Statistics and Computer Science, University of KwaZulu--Natal, Private Bag X54001, Durban 4000, South Africa}

\author{Bobur Turimov \orcidlink{0000-0003-1502-2053}}
\email{bturimov@astrin.uz}
\affiliation{University of Tashkent for Applied Sciences, Gavhar Str. 1, Tashkent, 100149, Uzbekistan}
\affiliation{Engineering School, Central Asian University, Milliy bog Str. 264, Tashkent, 111221, Uzbekistan}
\affiliation{Ulugh Beg Astronomical Institute, Astronomy Str. 33, Tashkent, 100052, Uzbekistan}

\author{Francisco Tello-Ortiz \orcidlink{0000-0002-7104-5746}}
\email{francisco.tello@ufrontera.cl}
\affiliation{Departamento de Ciencias Físicas, Universidad de La Frontera, Casilla 54-D, 4811186 Temuco, Chile.}

\author{Sulton Usanov\orcidlink{0009-0003-8351-708X}}
\email{sm.usanov@kiut.uz}
\affiliation{Kimyo International University in Tashkent, Usman Nasyr Str.156, Tashkent 100121, Uzbekistan}

\author{Farkhod Turaev}
\email{farhodjon9618@mail.ru}
\affiliation{Alfraganus University, Yukori Karakamish Str. 2a, Tashkent, 100190, Uzbekistan}
\date{\today}

\begin{abstract}
Gravitational-wave standard sirens provide a direct measurement of luminosity distance and therefore offer a new way to test gravity over cosmological scales. We use this idea to forecast the sensitivity of the Einstein Telescope (ET) to Lorentz violation in Bumblebee gravity, and we examine how the forecast changes when Type~Ia supernova information is added. A timelike Bumblebee vacuum expectation value can affect the cosmic expansion and, when it evolves with redshift, the propagation amplitude of gravitational waves. We study a constant-field case and an evolving-field case using mock ET catalogues with $10^3$ events together with a Pantheon+-like supernova sample. The supernova data substantially improve the background parameters: in the constant-field case the uncertainties in $H_0$ and $\Omega_m$ decrease by a factor of
about $4.4$, while in the evolving-field case they decrease by
factors of about $1.6$ and $6.2$, respectively. By contrast, the Lorentz-violating parameter $\ell_0$ remains prior dominated, and the evolution index $\beta$ is constrained only by the gravitational-wave sector. The best forecast precision, $\Delta\ell_0\simeq0.028$, is about $4.7\times10^{12}$ times weaker than the bound implied by GW170817. The principal result is therefore a quantified sensitivity gap rather than a forecast detection. We also express the prediction in the phenomenological $(\Xi,n)$ description of modified gravitational-wave propagation, allowing direct comparison with standard-siren studies of other gravity models.
\end{abstract}

\maketitle

\section{Introduction}
\label{sec:intro}

Lorentz invariance is built into both general relativity (GR) and the Standard Model, but it need not remain exact in every high-energy extension of known physics. Spontaneous Lorentz-symmetry breaking appears, for example, in string-inspired settings~\cite{Kostelecky:1989spontaneous}, and its low-energy consequences can be described systematically within the Standard-Model Extension~\cite{Colladay:1997cpt,Colladay:1998lorentzviolating,Kostelecky:2004gravity,Kostelecky:2009gravity}. This makes gravitational and cosmological observations useful laboratories for searching for small departures from local Lorentz symmetry.

Bumblebee gravity provides a simple realization of this idea. A vector field $B_\mu$ develops a nonzero vacuum expectation value (VEV), thereby selecting a preferred direction in spacetime~\cite{Bluhm2005}. Vacuum solutions and weak-field consequences of the model have been studied in detail~\cite{Bertolami:2005vacuum,Bailey:2006signals,Guiomar:2014astrophysical}. Its cosmological implications have also been explored, including homogeneous backgrounds, anisotropic extensions, and dark-energy applications~\cite{Capelo:2015cosmologicala,Maluf:2021bumblebee,Neves:2023kasner,Sarmah:2025anisotropic,Jesus:2019riccia,Siquieri:2026cosmic}. More recent analyses have connected Bumblebee cosmology with observational distance relations and perturbative stability~\cite{Zhu:2024bumblebee,Lai:2025stability,Gonzalez-Espinoza:2026cosmological}.

Gravitational waves (GWs) offer a particularly direct way to test this framework. Compact-binary mergers can act as ``standard sirens'', because the waveform gives an absolute luminosity distance without a conventional distance ladder~\cite{Schutz1986,HolzHughes2005}. In GR, the luminosity distance inferred from GWs follows the same background distance-redshift relation as electromagnetic (EM) signals. In modified gravity, however, the amplitude of a GW can acquire an additional propagation effect, so that the gravitational and electromagnetic luminosity distances need not coincide~\cite{Belgacem:2018gravitationalwave,Nishizawa:2018generalized}. This difference can be described in a model-independent way and has become a standard target for next-generation standard-siren forecasts~\cite{Belgacem:2018modified,Belgacem:2019testing}.

The multimessenger event GW170817/GRB~170817A already places an exceptionally strong restriction on any theory that changes the speed of tensor modes~\cite{Abbott:2017gravitational}. The near-equality of the GW and electromagnetic propagation speeds removed broad regions of parameter space in many modified-gravity models~\cite{Ezquiaga:2017dark,Baker:2017strong}. For the single-coupling Bumblebee model used here, the corresponding limit is $|\ellz|\lesssim6\times10^{-15}$~\cite{Lai:2025stability}. This bound is so small that the physically allowed cosmological signal is expected to be extremely weak. The relevant question is therefore not whether an illustrative large coupling can produce a visible departure from GR, but whether a large sample of future standard sirens can approach the existing GW170817 propagation-speed bound.

This paper addresses that question with mock Einstein Telescope data and a Pantheon+-like Type~Ia supernova sample. We consider two cases. In Case~I the timelike Bumblebee VEV is constant, so the observable modification enters through the background expansion. In Case~II the VEV evolves with redshift, which also changes the GW amplitude during propagation. The two cases allow us to separate what can be learned from the background expansion from what is genuinely specific to the GW sector. Our analysis follows the general forecasting logic used for standard sirens in other modified-gravity models, including the $F(Q)$ study of Ferreira et~al.~\cite{Ferreira:2022forecasting}.

The paper is organized as follows. Section~\ref{sec:model} introduces the Bumblebee cosmology and the GW luminosity distance used in the forecast. Section~\ref{sec:data_method} describes the mock data, likelihoods, priors, and numerical sampling. Section~\ref{sec:results} presents and discusses the results, including the comparison with the GW170817 bound. Section~\ref{sec:conclusions} summarizes the main conclusions. Technical comparisons with the parameterization of Zhu et~al. and with a LIGO-era mock catalogue are collected in the appendices. Throughout, $c=299\,792.458~\mathrm{km\,s^{-1}}$ and $H_0$ is expressed in $\mathrm{km\,s^{-1}\,Mpc^{-1}}$.

\section{Bumblebee cosmology and gravitational-wave propagation}
\label{sec:model}

\subsection{Cosmological background}
\label{sec:background}

We work with the single-coupling sector of Bumblebee gravity used in the perturbative analysis of Lai et~al.~\cite{Lai:2025stability}. The action is
\begin{widetext}
\begin{equation}
S=\int d^4x\sqrt{-g}\left[
\frac{R}{2\kappa}+\frac{\xi}{2\kappa}B^\mu B^\nu R_{\mu\nu}
-V(B_\mu B^\mu\pm b^2)-\frac14 B_{\mu\nu}B^{\mu\nu}
\right],
\label{eq:action}
\end{equation}
\end{widetext}
where $\kappa=8\pi G$, $B_{\mu\nu}=\partial_\mu B_\nu-\partial_\nu B_\mu$, and the potential $V$ drives the vector field toward a nonzero VEV. In a spatially homogeneous and isotropic FLRW background we take this VEV to be purely timelike,
\begin{equation}
\langle B_\mu\rangle=b_\mu=(b_t(t),0,0,0).
\end{equation}
The present amplitude is denoted by $b_{t,0}$, and we define the dimensionless combination controlling the observable effects as
\begin{equation}
\ellz\equiv\xi b_{t,0}^2.
\label{eq:ell0}
\end{equation}
The model can be written in terms of an effective dark-energy component with
\begin{align}
\rho_D&=\frac{\Lambda}{\kappa}+\frac{3\ellz H^2}{\kappa}+V,
\label{eq:rhoD}\\
p_D&=-\frac{\Lambda}{\kappa}-\frac{\ellz(3H^2+2\dot H)}{\kappa}-V,
\label{eq:pD}\\
w_D&=\frac{p_D}{\rho_D}=-1-\frac{2\ellz\dot H}{\kappa\rho_D}.
\label{eq:wD_general}
\end{align}
Thus, a nonzero timelike VEV can shift the effective equation of state away from $-1$. For the stable branch used below, $\ellz\le0$, the correction lies on the phantom side, $w_D\le-1$.

\subsection{Constant and evolving VEV}
\label{sec:cases}

For Case~I we take $b_t=b_{t,0}$ to be constant. The effective equation of state reduces to~\cite{Lai:2025stability}
\begin{equation}
w_D=-1-\frac{0.3|\ellz|}{1-\Om},
\label{eq:wD_caseI}
\end{equation}
and the normalized Hubble function is
\begin{equation}
E_{\rm I}(z)=\left[\Om(1+z)^3+(1-\Om)(1+z)^{3(1+w_D)}\right]^{1/2}.
\label{eq:EI}
\end{equation}
The GR/$\Lambda$CDM limit is recovered when $\ellz=0$. Even for the illustrative value $\ellz=-0.04$, Eq.~\eqref{eq:wD_caseI} gives only a modest shift, $w_D\simeq-1.017$ for $\Om=0.3$. At the physical GW170817 scale, the background departure is correspondingly much smaller.

In Case~II the VEV evolves as
\begin{equation}
b_t(z)=b_{t,0}(1+z)^{-\beta/2},
\end{equation}
where $\beta$ sets the rate of evolution. The background receives the correction
\begin{equation}
\Delta E^2(z)=\frac{3\ellz\beta}{1-\Om}
\int_0^z\frac{E_{\rm I}^2(z')}{(1+z')^{\beta+1}}\,dz'.
\label{eq:DeltaE2}
\end{equation}
At the fiducial values $(\Om,\ellz,\beta)=(0.30,-6\times10^{-15},2)$, this term is of relative order $10^{-14}$ over the redshift range relevant to the forecast. It is therefore negligible compared with the mock supernova and ET distance uncertainties. We consequently use Eq.~\eqref{eq:EI} as the background expansion in both cases and retain the distinctive Case~II effect in the propagation of the GW amplitude. A comparison with the alternative background parameterization of Zhu et~al.~\cite{Zhu:2024bumblebee} is given in Appendix~\ref{app:zhu}.

\subsection{GW propagation and luminosity distance}
\label{sec:gwprop}

Tensor perturbations in modified gravity can be written in the form~\cite{Belgacem:2018gravitationalwave,Nishizawa:2018generalized}
\begin{equation}
\tilde h_A''+(2+\nu)\mathcal H\tilde h_A'+\cT^2k^2\tilde h_A=0,
\label{eq:gw_eom}
\end{equation}
where $\mathcal H=aH$, $\nu$ describes an additional friction term, and $\cT$ is the tensor propagation speed. The polarization content of Bumblebee gravity has been studied explicitly in Ref.~\cite{Liang:2022polarizations}. For the two cases considered here, the tensor kinetic coefficient is
\begin{align}
\GT^{\rm (I)}&=1-\ellz,
\label{eq:GT_const}\\
\GT^{\rm (II)}(z)&=1-\ellz(1+z)^{-\beta}.
\label{eq:GT_case2}
\end{align}
The second expression is redshift dependent and therefore produces an additional damping of the GW amplitude. The corresponding friction term is
\begin{equation}
\nu(z)=-\frac{\ellz\beta(1+z)^{-\beta}}
{1-\ellz(1+z)^{-\beta}}.
\label{eq:nu_case2}
\end{equation}
For a constant VEV, $\beta=0$ and this additional friction vanishes.

The observable consequence is most simply expressed through the GW/EM luminosity-distance ratio,
\begin{equation}
\frac{\dLGW(z)}{\dLEM(z)}=
\sqrt{\frac{\GT(0)}{\GT(z)}}.
\label{eq:ratio_general}
\end{equation}
The electromagnetic luminosity distance is
\begin{equation}
\dLEM(z)=\frac{c}{H_0}(1+z)\int_0^z\frac{dz'}{E_{\rm I}(z';\Om,\ellz)}.
\label{eq:dLEM}
\end{equation}
In Case~I, $\GT$ is constant and therefore
\begin{equation}
\dLGW=\dLEM.
\label{eq:caseI_distance}
\end{equation}
The Bumblebee signature then enters only through the background expansion. In Case~II,
\begin{equation}
\frac{\dLGW}{\dLEM}\bigg|_{\rm II}=
\sqrt{\frac{1-\ellz}{1-\ellz(1+z)^{-\beta}}},
\label{eq:ratio_case2}
\end{equation}
which, to first order in $\ellz$, becomes
\begin{equation}
\frac{\dLGW}{\dLEM}\simeq1+\frac{|\ellz|}{2}
\left[1-(1+z)^{-\beta}\right]
\quad (\ellz<0,\ \beta>0).
\label{eq:ratio_approx}
\end{equation}
Thus an evolving VEV makes a distant GW source appear slightly farther away than the same source would appear through an EM distance indicator. For $\ellz=-0.04$ and $\beta=2$, the effect is about $1.5\%$ at $z=1$. This large coupling is used only to display the shape of the effect. At $|\ellz|\sim10^{-15}$ the difference is itself of order $10^{-15}$.

It is convenient to combine Eqs.~\eqref{eq:dLEM} and~\eqref{eq:ratio_case2} as
\begin{equation}
\dLGW(z)=\dLEM(z)\,\mathcal R(z;\ellz,\beta),
\label{eq:dLGW_full}
\end{equation}
with
\begin{equation}
\mathcal R(z)=
\begin{cases}
1, & \text{Case~I},\\[4pt]
\displaystyle\sqrt{\dfrac{1-\ellz}{1-\ellz(1+z)^{-\beta}}}, & \text{Case~II}.
\end{cases}
\label{eq:Rfactor}
\end{equation}
This factorization is useful because it separates the background information, which is also accessible to supernovae, from the propagation factor that is specific to GWs.

\subsection{Viability and phenomenological mapping}
\label{sec:viability}

The tensor speed in the single-coupling model is
\begin{equation}
\cT^2=\frac{1}{1-\ellz}.
\label{eq:cT2}
\end{equation}
The GW170817/GRB~170817A constraint on the propagation speed therefore requires $|\ellz|\lesssim6\times10^{-15}$~\cite{Abbott:2017gravitational,Lai:2025stability}. An independent, weaker constraint on the same coupling has been obtained from big-bang nucleosynthesis and gravitational baryogenesis~\cite{Khodadi:2023constraining}. We also impose the perturbative viability conditions derived in Ref.~\cite{Lai:2025stability},
\begin{align}
1-\ellz&>0, & \cT^2&>0, & \ellz&\le0,
\label{eq:stability}
\end{align}
with $\GT^{\rm(II)}(z)>0$ required throughout the Case~II redshift range. These conditions are implemented as hard priors in the statistical analysis.

We restrict the forecast to the single-coupling theory. The two-coupling Bumblebee extensions analyzed in Refs.~\cite{VanDeBruck:2026nogo,Nilsson:2025bumblebee} do not provide a viable cancellation that removes the tensor-speed restriction, and their perturbative structure introduces further difficulties. This is why no relaxed $\mathcal{O}(10^{-1})$ physical prior on $\ell_0$ is used here, even though a wider numerical interval is sampled to assess the sensitivity achievable by the forecast.

For comparison with the wider modified-gravity literature, we also use the phenomenological relation~\cite{Belgacem:2018modified}
\begin{equation}
\frac{\dLGW}{\dLEM}=\Xi+(1-\Xi)(1+z)^{-n}.
\label{eq:belgacem}
\end{equation}
At first order in $\ellz$, Eqs.~\eqref{eq:ratio_approx} and~\eqref{eq:belgacem} give
\begin{equation}
\Xi=1-\frac{\ellz}{2},\qquad n=\beta.
\label{eq:Xi_n_map}
\end{equation}
At the fiducial point, $\Xi-1=3\times10^{-15}$, so the physically allowed departure from the GR value $\Xi=1$ is far below percent-level distance measurements.

\section{Data and methodology}
\label{sec:data_method}

\subsection{Mock standard sirens and supernovae}
\label{sec:catalogues}

The main GW catalogue is designed to represent an idealized ET standard-siren forecast. The Einstein Telescope is a third-generation ground-based detector expected to observe compact-binary mergers to substantially greater distances and with better precision than current instruments~\cite{Punturo:2010einstein}. We generate $N_{\rm GW}=1000$ events with redshifts drawn from a volume-weighted distribution $p(z)\propto z^2$ up to $z_{\max}=2$. Each event is assigned a fractional distance uncertainty drawn uniformly from $5\%$ to $15\%$. The observed distance is then obtained from a Gaussian distribution centered on the theoretical $\dLGW(z_i)$ evaluated at
\begin{equation}
(H_0,\Om,\ellz,\beta)=(67.64,0.30,-6\times10^{-15},2.0).
\label{eq:fiducial}
\end{equation}
The injected Lorentz-violating coupling therefore lies at the edge of the GW170817 bound. A smaller LIGO-era catalogue with $N=100$, $z<0.5$, and $10$--$30\%$ distance errors is used only as a detector-generation comparison.

The electromagnetic information is represented by a Pantheon+-like mock Type~Ia supernova catalogue. Pantheon+ contains 1701 supernova light curves over a broad redshift range~\cite{Scolnic:2022pantheon}. We therefore generate $N_{\rm SN}=1701$ mock events with $z\in[0.01,2.3]$, using a log-uniform redshift distribution and magnitude uncertainties drawn from $0.08$--$0.15$~mag. The mock apparent magnitudes are generated with $M_B=-19.25$~mag from
\begin{equation}
m_{B,i}=M_B+5\log_{10}\!\left(\frac{c/H_0}{\mathrm{Mpc}}\right)+25
+5\log_{10}D_{L,i}^{\rm dimless}+\epsilon_i,
\label{eq:mock_snia}
\end{equation}
where $D_{L,i}^{\rm dimless}=(1+z_i)\int_0^{z_i}dz'/E_{\rm I}(z')$ and $\epsilon_i\sim\mathcal N(0,\sigma_{m,i}^2)$. Because the absolute magnitude is marginalized in the likelihood, its fiducial value does not enter the recovered cosmological constraints.

\subsection{Likelihoods and priors}
\label{sec:likelihoods}

The parameter vectors are
\begin{equation}
\bm\theta_{\rm I}=\{H_0,\Om,\ellz\},\qquad
\bm\theta_{\rm II}=\{H_0,\Om,\ellz,\beta\}.
\end{equation}
We use the flat priors listed in Table~\ref{tab:priors}. The interval $\ellz\in[-0.1,0]$ is intentionally much wider than the physical GW170817 limit. It is a numerical forecast interval, not an alternative physical bound. Sampling a wide interval lets the posterior show directly whether ET-like data can approach the existing constraint.

\begin{table}[t]
\centering
\caption{Flat prior ranges used in the MCMC forecast. The physical fiducial value is $\ellz=-6\times10^{-15}$; the wider numerical interval for $\ellz$ is used to measure the achievable sensitivity.}
\label{tab:priors}
\begin{ruledtabular}
\begin{tabular}{lcc}
Parameter & Prior range & Case \\
\midrule
$H_0$ [$\mathrm{km\,s^{-1}\,Mpc^{-1}}$] & $[50,100]$ & I, II \\
$\Om$ & $[0.1,0.9]$ & I, II \\
$\ellz$ & $[-0.1,0]$ & I, II \\
$\beta$ & $[-3.0,5.0]$ & II \\
\end{tabular}
\end{ruledtabular}
\end{table}

\begin{figure*}[!htbp]
\centering
\includegraphics[width=\textwidth]{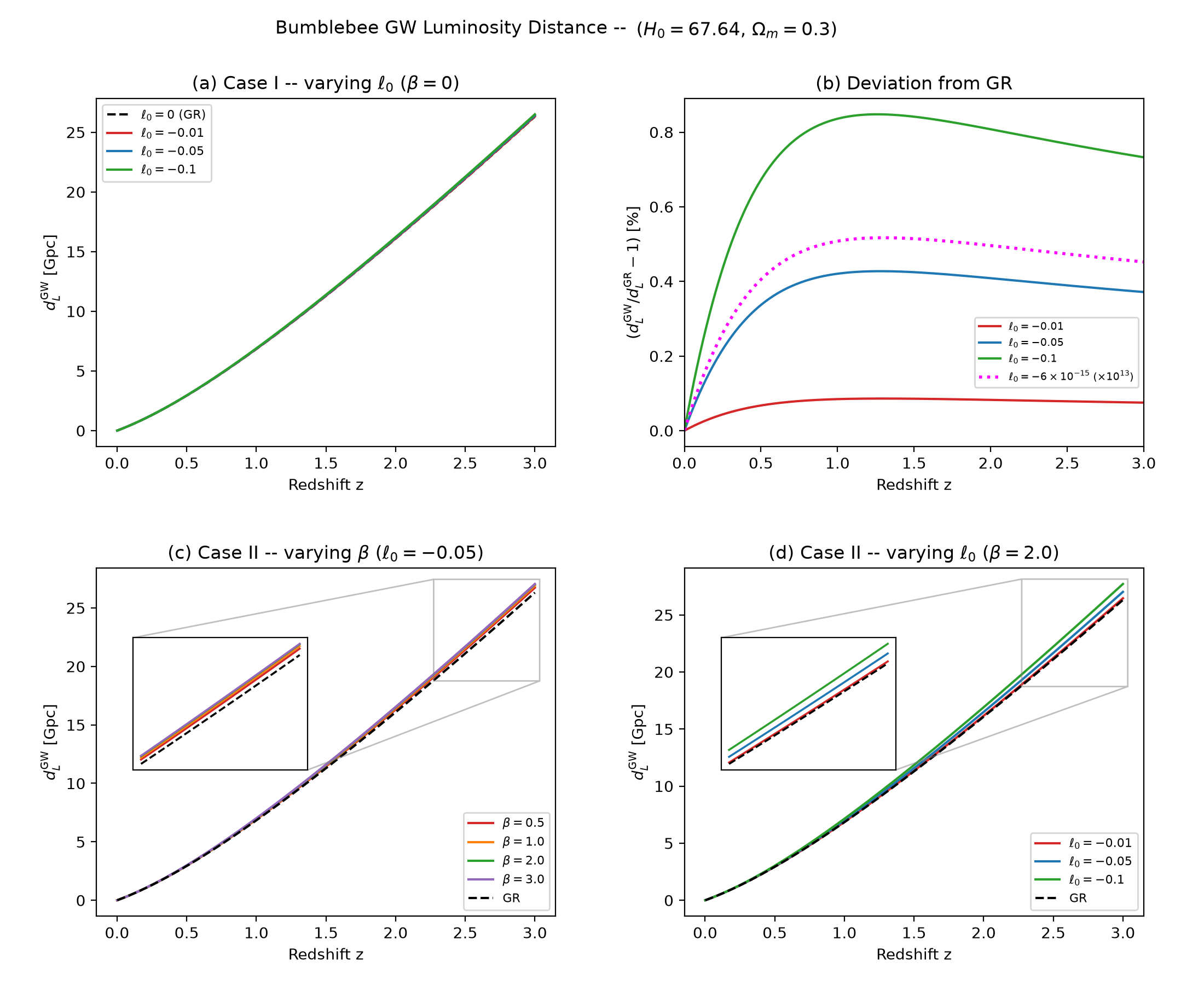}
\caption{GW luminosity distance in Bumblebee gravity for $H_0=67.64\;\mathrm{km\,s^{-1}\,Mpc^{-1}}$ and $\Om=0.3$. Panels (a), (c), and (d) use enlarged illustrative values of $\ellz$ to show the shape of the modification; these values are not physically allowed by the GW170817 bound. Panel (a) shows Case~I, panel (b) shows the relative deviation from GR, panel (c) varies $\beta$ at fixed $\ellz=-0.05$, and panel (d) varies $\ellz$ at fixed $\beta=2$. In panel (b) the physical bound-edge curve is rescaled by $10^{13}$ for visibility.}
\label{fig:dL_curves}
\end{figure*}

For the standard sirens we adopt a Gaussian likelihood,
\begin{multline}
\ln\mathcal L_{\rm GW}(\bm\theta)=
-\frac{N_{\rm GW}}2\ln(2\pi)-\sum_{i=1}^{N_{\rm GW}}\ln\sigma_i\\
-\frac12\sum_{i=1}^{N_{\rm GW}}
\left[\frac{d_{L,i}^{\rm obs}-\dLGW(z_i;\bm\theta)}{\sigma_i}\right]^2.
\label{eq:ll_gw}
\end{multline}
For the supernovae, we marginalize analytically over the absolute magnitude. We follow the forecasting procedure of Ferreira et~al.~\cite{Ferreira:2022forecasting}, whose public implementation is also used as a methodological reference~\cite{Ferreira:2022repo}. That work was developed in the context of $F(Q)$ cosmology~\cite{Jimenez:2018coincident,Jimenez:2020cosmology}, a class of models that has also been confronted with cosmological observations through independent probes~\cite{Ayuso:2021observational,Barros:2020testing}. Defining
\begin{equation}
\delta_i=m_{B,i}-5\log_{10}\!\left[(1+z_i)\int_0^{z_i}\frac{dz'}{E_{\rm I}(z')}\right],
\label{eq:delta_i}
\end{equation}
the marginalized SNIa likelihood is written as
\begin{equation}
\ln\mathcal L_{\rm SNIa}=-A+\frac{B^2}{C},
\label{eq:ll_snia}
\end{equation}
with
\begin{equation}
A=\sum_i\frac{\delta_i^2}{\sigma_{m,i}^2},\qquad
B=\sum_i\frac{\delta_i}{\sigma_{m,i}^2},\qquad
C=\sum_i\frac1{\sigma_{m,i}^2}.
\label{eq:ABC}
\end{equation}
The joint posterior is therefore
\begin{equation}
\ln p(\bm\theta|\mathrm{data})=\ln\pi(\bm\theta)
+\ln\mathcal L_{\rm GW}+\ln\mathcal L_{\rm SNIa}.
\label{eq:posterior}
\end{equation}
The distinction between the two datasets is important. Supernovae measure $\dLEM$ and constrain the background expansion. They do not measure the propagation factor $\mathcal R(z)$ and therefore do not directly constrain $\beta$. Standard sirens, on the other hand, are sensitive to both pieces of Eq.~\eqref{eq:dLGW_full}.

\subsection{Sampling and numerical evaluation}
\label{sec:sampling}

We sample the posterior with the affine-invariant \texttt{emcee} ensemble sampler~\cite{Foreman-Mackey:2013emcee}. Each run uses 64 walkers and 3000 steps. The first 1000 steps are discarded as burn-in and the remaining chains are thinned by a factor of six, leaving approximately $2.1\times10^4$ posterior samples. The mean acceptance fraction lies between $0.40$ and $0.61$ for the runs reported below.

The luminosity-distance integral must be evaluated many times inside the MCMC. We therefore evaluate $1/E_{\rm I}(z)$ on a 600-point grid, form the cumulative integral with the trapezoidal rule, and interpolate it to the catalogue redshifts. Direct comparisons with per-event Gaussian quadrature give relative differences below $0.01\%$ across the sampled parameter region. This numerical shortcut changes the computational cost, not the forecast model.

\section{Results and discussion}
\label{sec:results}

\subsection{How Bumblebee gravity changes the luminosity distance}
\label{sec:dL_phenomenology}

Figure~\ref{fig:dL_curves} first shows the qualitative form of the signal. In Case~I the GW and EM luminosity distances are identical, and the departure from GR comes only from the small change in the background expansion. The plotted values $\ellz=-0.01,-0.05$, and $-0.1$ are intentionally far larger than the physical bound so that the shape can be seen. For $\ellz=-0.1$ the relative difference reaches about $0.73\%$ at $z=3$, while for $\ellz=-0.04$ it is about $0.33\%$ at $z=2$. At the physical value $\ellz=-6\times10^{-15}$ the corresponding departure is of order $10^{-13}\%$ and is visually indistinguishable from the GR curve.

Case~II adds the propagation factor $\mathcal R(z)$. At fixed $\ellz$, increasing $\beta$ makes the ratio approach its high-redshift limit more quickly. At fixed $\beta$, increasing $|\ellz|$ enlarges both the background and propagation effects. The figure is therefore useful for understanding the direction of the modification, but it also makes the scale problem clear: the coupling values needed for a visible percent-level separation are many orders of magnitude above the GW170817 limit.

\subsection{From a LIGO-era catalogue to the Einstein Telescope}
\label{sec:ligo_vs_et}

\begin{figure}[!htbp]
\centering
\includegraphics[width=\columnwidth]{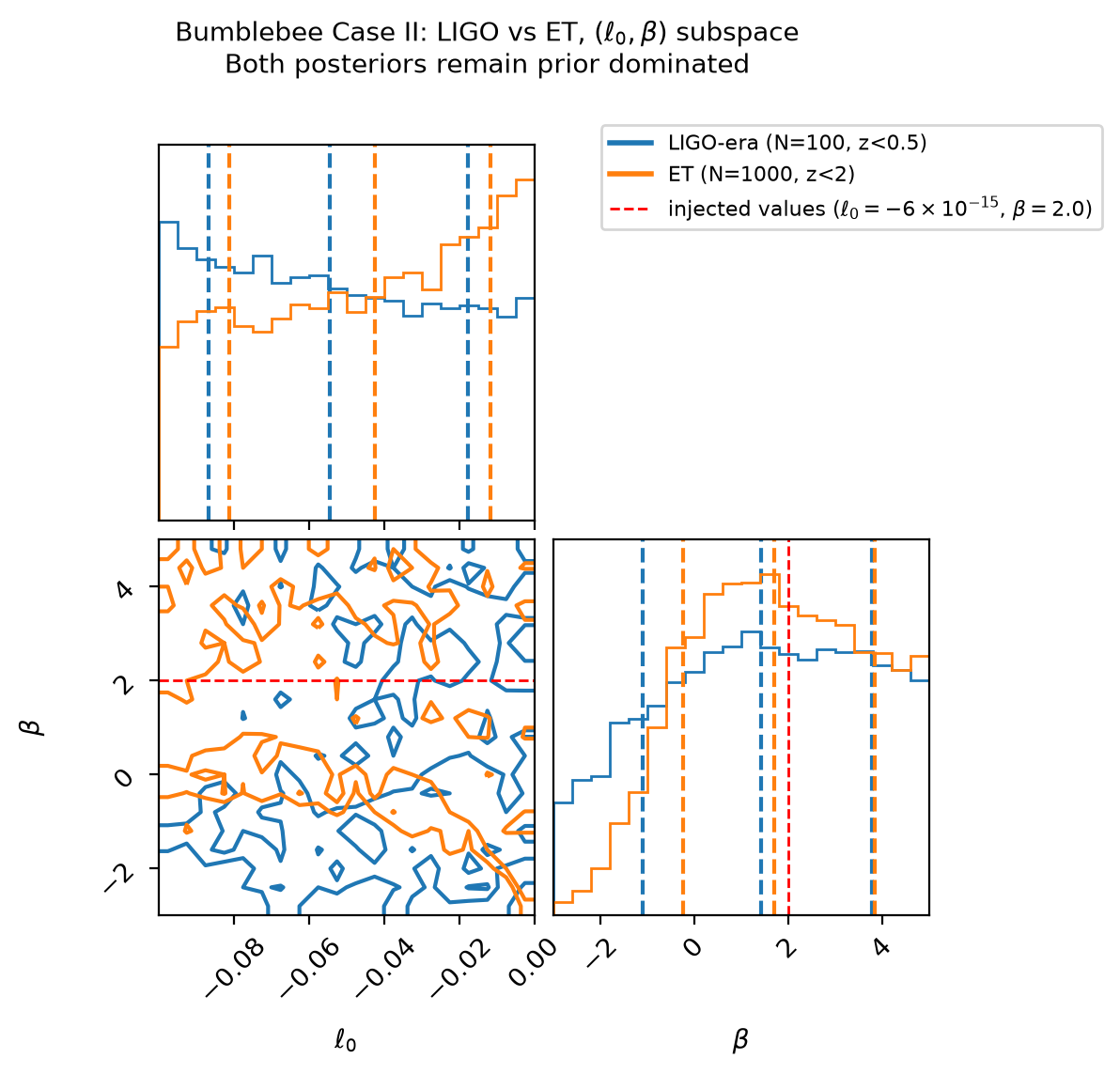}
\caption{Case~II GW-only posterior in the $(\ellz,\beta)$ plane for the LIGO-era (blue) and ET (orange) mock catalogues. The red dashed lines mark the injected values $\ellz=-6\times10^{-15}$ and $\beta=2$. ET narrows the cosmological parameter space, but the $\ellz$ marginal remains broad and prior dominated.}
\label{fig:ligo_vs_et}
\end{figure}

Figure~\ref{fig:ligo_vs_et} compares Case~II GW-only posteriors obtained from the LIGO-era and ET mock catalogues. ET improves the background constraints because it contains more events, reaches higher redshift, and assigns smaller distance errors. The Lorentz-violating parameter behaves differently. Its $68\%$ interval remains broad in both cases, approximately $[-0.087,-0.018]$ for the LIGO-era mock catalogue and $[-0.081,-0.012]$ for ET. The widths are therefore set mainly by the adopted prior rather than by the data. The full four-parameter comparison is given in Appendix~\ref{app:ligo_et}.

\subsection{Case~I: constant Bumblebee VEV}
\label{sec:case1_results}

\begin{figure*}[!htbp]
\centering
\includegraphics[width=0.85\textwidth]{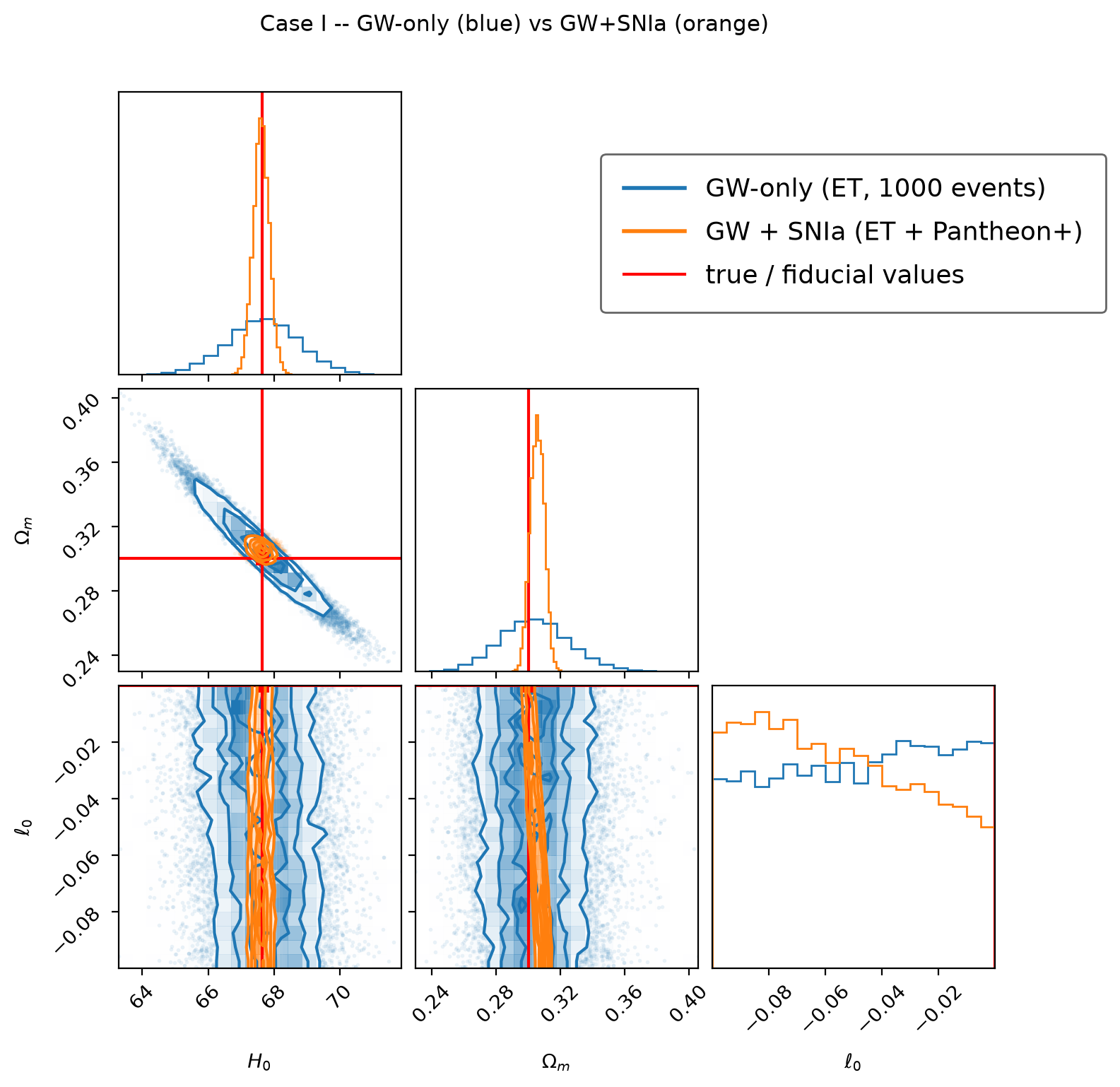}
\caption{Case~I posterior distributions for $\{H_0,\Om,\ellz\}$ from ET standard sirens alone (blue) and from the joint GW+SNIa analysis (orange). Red lines mark the injected values. The supernova sample sharply narrows the background parameters, while the $\ellz$ posterior remains essentially unchanged.}
\label{fig:case1_comparison}
\end{figure*}

The Case~I comparison is shown in Fig.~\ref{fig:case1_comparison} and Table~\ref{tab:case1_results}. With ET standard sirens alone, the recovered uncertainties are $\Delta H_0\simeq1.11$ and $\Delta\Om\simeq0.022$. The posterior is summarized by $\ellz=-0.049\pm0.029$. This value should not be interpreted as evidence for Lorentz violation: the injected coupling is effectively zero on the scale of the posterior, so the broad result simply reflects the weak sensitivity of the forecast.

Adding the Pantheon+-like supernova sample changes the background constraints substantially. The uncertainty in $H_0$ falls to about $0.25$ and the uncertainty in $\Om$ to about $0.005$, an improvement of $4.4\times$ in each case. The uncertainty in $\ellz$ is essentially unchanged. This pattern has a simple origin. In Case~I both standard sirens and supernovae follow the same luminosity-distance relation; the supernovae sharpen the background expansion, but the effect of $\ellz$ on that background is too small at the physical fiducial value to be isolated.

\begin{table}[!htbp]
\centering
\caption{Case~I posterior constraints (median $\pm1\sigma$). The injected values are $(H_0,\Om,\ellz)=(67.64,0.30,-6\times10^{-15})$.}
\label{tab:case1_results}
\begin{ruledtabular}
\begin{tabular}{lccc}
Parameter & GW-only & GW+SNIa & Improvement \\
\midrule
$H_0$ & $67.71\pm1.11$ & $67.60\pm0.25$ & $4.4\times$ \\
$\Om$ & $0.302\pm0.022$ & $0.306\pm0.005$ & $4.4\times$ \\
$\ellz$ & $-0.049\pm0.029$ & $-0.058\pm0.028$ & $1.0\times$ \\
\end{tabular}
\end{ruledtabular}
\end{table}

\subsection{Case~II: evolving Bumblebee VEV}
\label{sec:case2_results}

\begin{figure*}[!htbp]
\centering
\includegraphics[width=0.85\textwidth]{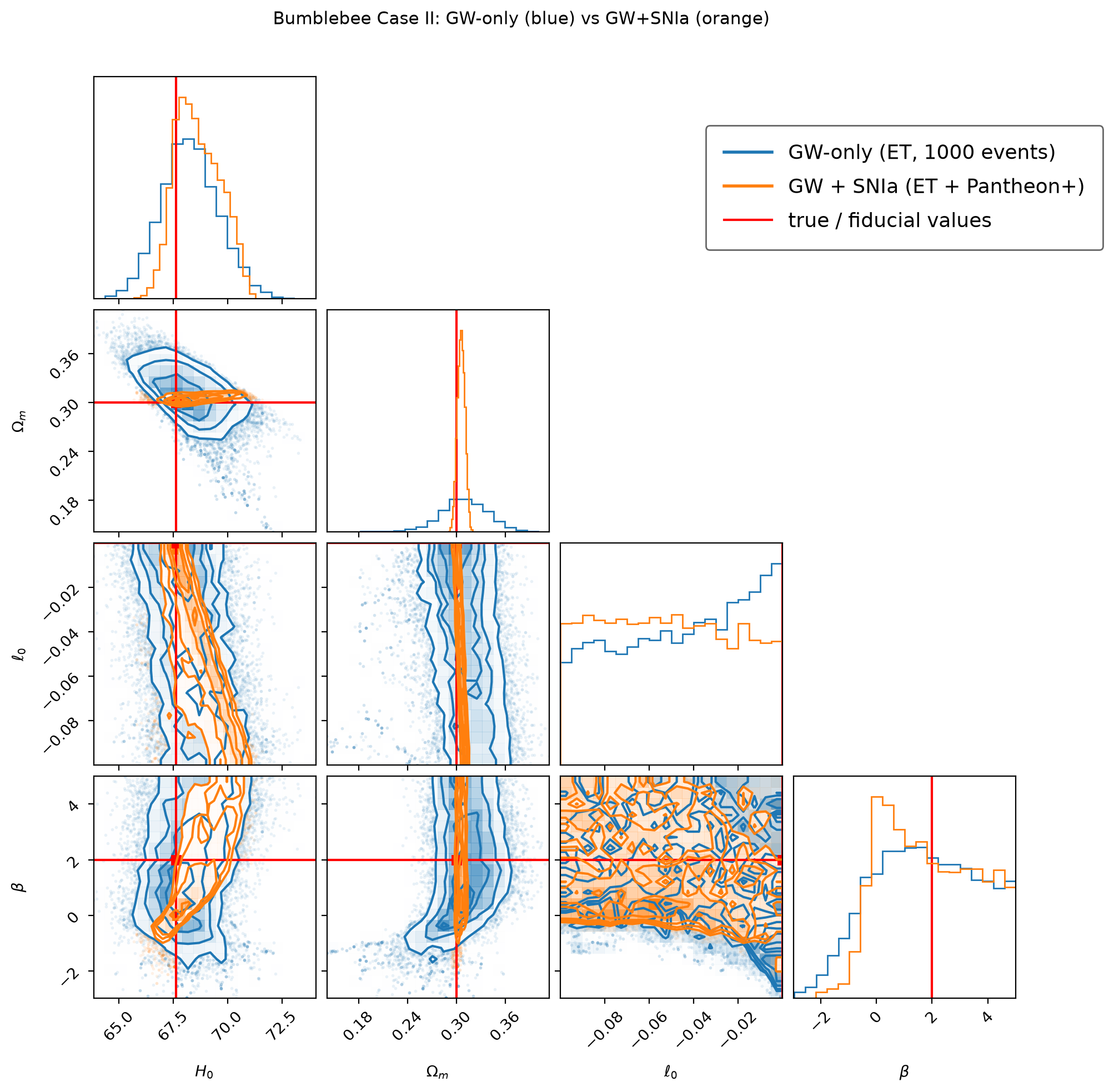}
\caption{Case~II posterior distributions for $\{H_0,\Om,\ellz,\beta\}$ from ET standard sirens alone (blue) and from the joint GW+SNIa analysis (orange). Red lines mark the injected values. The supernova sample mainly improves $H_0$ and $\Om$; the Lorentz-violating parameters remain controlled by the GW sector.}
\label{fig:case2_comparison}
\end{figure*}

Case~II introduces the additional index $\beta$, which controls the redshift dependence of the GW propagation factor. Figure~\ref{fig:case2_comparison} and Table~\ref{tab:case2_results} show that the supernova sample again improves the background parameters. The uncertainty in $\Om$ decreases from $0.031$ to $0.005$, a factor of $6.2$, while that in $H_0$ decreases from $1.35$ to $0.85$, a factor of $1.6$. The smaller improvement in $H_0$ than in Case~I reflects the extra freedom introduced by $\beta$.

Neither Lorentz-violating parameter benefits in the same way. The width of the $\ellz$ posterior remains about $0.03$, and the uncertainty in $\beta$ is $1.94$ for GW-only and $2.15$ for GW+SNIa. The small numerical widening should not be overinterpreted. Supernovae have no direct access to $\beta$, because $\beta$ appears only in $R(z)$, whereas SNIa measure $d_L^{\rm EM}$. This is the clearest expression of the complementarity in the model: electromagnetic distances determine the background more accurately, while only standard sirens probe the redshift-dependent GW propagation term.

\begin{table}[!htbp]
\centering
\caption{Case~II posterior constraints (median $\pm1\sigma$). The injected values are $(H_0,\Om,\ellz,\beta)=(67.64,0.30,-6\times10^{-15},2.0)$.}
\label{tab:case2_results}
\begin{ruledtabular}
\begin{tabular}{lccc}
Parameter & GW-only & GW+SNIa & Improvement \\
\midrule
$H_0$ & $68.07\pm1.35$ & $67.75\pm0.85$ & $1.6\times$ \\
$\Om$ & $0.308\pm0.031$ & $0.301\pm0.005$ & $6.2\times$ \\
$\ellz$ & $-0.029\pm0.030$ & $-0.0018\pm0.029$ & $1.0\times$ \\
$\beta$ & $1.70\pm1.94$ & $1.79\pm2.15$ & $1.0\times$ \\
\end{tabular}
\end{ruledtabular}
\end{table}

\subsection{What the forecast can test}
\label{sec:sensitivity}

The most important result emerges when the recovered width of $\ellz$ is compared with the existing propagation-speed bound. Across the analyses, the $1\sigma$ width is close to $0.029$, which is also the standard deviation expected from an almost flat prior over $[-0.1,0]$, namely $0.1/\sqrt{12}\simeq0.029$. The best value occurs in the Case~I joint run, $\Delta\ellz\simeq0.028$. Relative to $|\ellz|\lesssim6\times10^{-15}$, this gives
\begin{equation}
\frac{\Delta\ellz^{\rm best}}{\ellz^{\rm bound}}
=\frac{0.028}{6\times10^{-15}}
\simeq4.7\times10^{12}.
\label{eq:sensitivity_gap}
\end{equation}
The forecast is therefore more than twelve orders of magnitude away from testing or improving the GW170817 bound. This is not a failure of the MCMC to recover an injected signal; the injected signal is simply much smaller than the sensitivity afforded by the mock distance measurements.

The same point is visible in the phenomenological parameters of Eq.~\eqref{eq:belgacem}. The Case~II GW-only posterior corresponds to
\begin{equation}
\Xi=1.0145\pm0.015,\qquad n=1.70\pm1.94.
\end{equation}
The percent-level width in $\Xi$ is fully consistent with GR, but it is not a measurement of the physical Bumblebee departure. At the injected coupling, $\Xi-1$ is only $3\times10^{-15}$. In other words, the model can be translated cleanly into the standard modified-propagation language, but the physically allowed deviation remains far below the forecast resolution~\cite{Belgacem:2018gravitationalwave}.

This sensitivity gap also helps place the analysis in the context of recent Bumblebee cosmology. Zhu et~al.~\cite{Zhu:2024bumblebee} studied distance- and time-redshift relations using a different background parameterization, while Lai et~al.~\cite{Lai:2025stability} derived the perturbative stability conditions and the present GW propagation constraint. Our forecast asks a complementary question: given that bound, how much additional leverage can future cosmological distance measurements provide? For the catalogue assumed here, the answer is that ET and SNIa strongly improve standard background parameters but do not approach the existing constraint on $\ellz$.

Several limitations should be kept in mind. The catalogues use simplified Gaussian distance and magnitude errors and omit correlated systematics, selection effects, calibration uncertainties, lensing reconstruction, and peculiar-velocity modelling beyond the adopted error prescription. The ET forecast also assumes that all $10^3$ events have identified EM counterparts and therefore measured redshifts. These choices make the forecast deliberately idealized. They can change the detailed parameter errors, but they do not remove the basic scale separation between an $\mathcal O(10^{-2})$ posterior width and an $\mathcal O(10^{-15})$ physical bound.

Other cosmological probes can test different aspects of the same theory. CMB calculations already show strong sensitivity to Bumblebee-driven departures from $\Lambda$CDM away from the very small GW170817-compatible region~\cite{Xu:2026bumblebee}. Primordial gravitational waves provide an independent early-Universe channel~\cite{Khodadi:2025primordial}. Future work could therefore combine standard sirens with CMB, BAO, or high-redshift space-based GW observations. Such combinations are useful not because the present ET forecast suggests a detectable $\ellz$, but because they test complementary sectors and epochs of Lorentz-violating cosmology.

\section{Conclusions}
\label{sec:conclusions}

We have studied whether future standard-siren distances can provide a useful cosmological test of Lorentz violation in Bumblebee gravity. The calculation was organized around two simple possibilities for the timelike Bumblebee VEV. When the VEV is constant, the modification enters through the background expansion and the GW and EM luminosity distances remain equal. When the VEV evolves, the GW amplitude acquires an additional redshift-dependent propagation factor, so standard sirens contain information that is absent from Type~Ia supernova distances.

The mock-data analysis shows a clear division of roles between the two probes. Pantheon+-like supernovae substantially improve $H_0$ and $\Omega_m$, reducing their uncertainties by $4.4\times$ in Case~I and by $1.6\times$ and $6.2\times$, respectively, in Case~II. They do not comparably improve $\ellz$, and they do not constrain $\beta$ because that parameter acts only through GW propagation. Standard sirens therefore supply the distinctive tensor-sector information, while supernovae mainly anchor the background cosmology.

The central quantitative result is the sensitivity gap. The narrowest recovered uncertainty is $\Delta\ellz\simeq0.028$, whereas GW170817 requires $|\ellz|\lesssim6\times10^{-15}$. The forecast is consequently weaker by a factor of about $4.7\times10^{12}$. At the physical fiducial value, neither the background correction nor the GW/EM distance difference can be resolved by the ET catalogue assumed here. The broad posterior on $\ellz$ should therefore be interpreted as a non-detection with prior-dominated sensitivity, not as a measurement away from GR.

This result does not diminish the value of standard sirens as a test of Bumblebee gravity. It identifies precisely what they can and cannot do in this model. They provide a clean cosmological probe of the tensor sector and a direct way to compare gravitational and electromagnetic distance measures, but the current propagation-speed bound is already far stronger than the reach of the forecast considered here. A natural next step is therefore to combine standard sirens with probes that respond to different aspects of Bumblebee cosmology, rather than to expect ET distances alone to improve the GW170817 constraint.

\appendix

\section{Relation to the Zhu et~al. background parameterization}
\label{app:zhu}

Zhu et~al.~\cite{Zhu:2024bumblebee} write the Bumblebee background in terms of
\begin{multline}
E_{\rm Zhu}(z)=\Big[\Omega_{V_1}
+(1-\Omega_{V_1}-\Omega_{K_0})(1+z)^{2-\alpha}\\
+\Omega_{K_0}(1+z)^2\Big]^{1/2},
\label{eq:EZhu_app}
\end{multline}
with
\begin{equation}
\Omega_{V_1}=\frac{2\tilde V_1}{\xi_1+4\xi_2},\qquad
\alpha=\frac{-4\xi_2}{\xi_1+2\xi_2},
\label{eq:Zhu_composite_app}
\end{equation}
and $\tilde V_1=8\pi V_1/(3H_0^2)$. In the flat limit, $\Omega_{K_0}=0$,
\begin{equation}
E_{\rm Zhu}^2(z)=\Omega_{V_1}+(1-\Omega_{V_1})(1+z)^{2-\alpha}.
\end{equation}
The $\Lambda$CDM limit of our Case~I expression is obtained by identifying $1-\Omega_{V_1}=\Om$, $\Omega_{V_1}=1-\Om$, and $2-\alpha=3$. For nonzero $\ellz$, the two descriptions use different functional forms for the departure from $\Lambda$CDM, so the correspondence is parameter dependent rather than a one-to-one identification at all redshifts. This appendix is included only to make the relation between the two conventions explicit.

\section{LIGO-era and ET mock-catalogue comparison}
\label{app:ligo_et}

Table~\ref{tab:ligo_et} gives the full Case~II GW-only comparison behind Fig.~\ref{fig:ligo_vs_et}. ET substantially improves $H_0$ and $\Om$, but the $\ellz$ uncertainty remains $\simeq0.029$ for both mock catalogues. The evolution index $\beta$ improves only modestly. The comparison reinforces the main result that increasing the number and reach of standard sirens helps the ordinary cosmological parameters much more than it helps a coupling whose physically allowed effect is already of order $10^{-15}$.

\onecolumngrid
\begin{table}[H]
\centering
\caption{Case~II GW-only posterior comparison between the LIGO-era and ET mock catalogues.}
\label{tab:ligo_et}
\begin{ruledtabular}
\begin{tabular}{lccc}
Parameter
& LIGO
& ET
& Improvement \\

& {\scriptsize $(N=100,\ z<0.5)$}
& {\scriptsize $(N=1000,\ z<2)$}
& \\

\midrule
$H_0$
& $68.84\pm3.27$
& $68.21\pm1.32$
& $2.5\times$ \\

$\Omega_m$
& $0.279\pm0.175$
& $0.309\pm0.032$
& $5.5\times$ \\

$\ell_0$
& $-0.054\pm0.029$
& $-0.042\pm0.029$
& $1.0\times$ \\

$\beta$
& $1.42\pm2.14$
& $1.71\pm1.83$
& $1.2\times$ \\
\end{tabular}
\end{ruledtabular}
\end{table}
\twocolumngrid

\bibliographystyle{apsrev4-2}
\bibliography{references}

@article{Abbott:2017gravitational,
  title        = {Gravitational Waves and Gamma-Rays from a Binary Neutron
                  Star Merger: {GW}170817 and {GRB} 170817{A}},
  author       = {Abbott, B. P. and others},
  collaboration = {LIGO Scientific Collaboration and Virgo Collaboration
                  and Fermi Gamma-ray Burst Monitor and INTEGRAL},
  year         = 2017,
  journal      = {Astrophys. J. Lett.},
  volume       = {848},
  number       = {2},
  pages        = {L13},
  doi          = {10.3847/2041-8213/aa920c}
}

@article{Baker:2017strong,
  title   = {Strong Constraints on Cosmological Gravity from {GW}170817
             and {GRB} 170817{A}},
  author  = {Baker, T. and Bellini, E. and Ferreira, P. G. and Lagos, M.
             and Noller, J. and Sawicki, I.},
  year    = 2017,
  journal = {Phys. Rev. Lett.},
  volume  = {119},
  number  = {25},
  pages   = {251301},
  doi     = {10.1103/PhysRevLett.119.251301}
}

@article{Belgacem:2018gravitationalwave,
  title   = {Gravitational-Wave Luminosity Distance in Modified Gravity
             Theories},
  author  = {Belgacem, Enis and Dirian, Yves and Foffa, Stefano and
             Maggiore, Michele},
  year    = 2018,
  journal = {Phys. Rev. D},
  volume  = {97},
  number  = {10},
  pages   = {104066},
  doi     = {10.1103/PhysRevD.97.104066}
}

@article{Belgacem:2018modified,
  title   = {Modified Gravitational-Wave Propagation and Standard Sirens},
  author  = {Belgacem, Enis and Dirian, Yves and Foffa, Stefano and
             Maggiore, Michele},
  year    = 2018,
  journal = {Phys. Rev. D},
  volume  = {98},
  number  = {2},
  pages   = {023510},
  doi     = {10.1103/PhysRevD.98.023510}
}

@article{Belgacem:2019testing,
  title   = {Testing Modified Gravity at Cosmological Distances with
             {LISA} Standard Sirens},
  author  = {Belgacem, Enis and Calcagni, Gianluca and Crisostomi, Marco
             and Dalang, Charles and Dirian, Yves and Ezquiaga, Jose
             Mar{\'i}a and Fasiello, Matteo and Foffa, Stefano and Ganz,
             Alexander and Garc{\'i}a-Bellido, Juan and Lombriser, Lucas
             and Maggiore, Michele and Tamanini, Nicola and Tasinato,
             Gianmassimo and Zumalac{\'a}rregui, Miguel and Barausse,
             Enrico and Bartolo, Nicola and Bertacca, Daniele and Klein,
             Antoine and Matarrese, Sabino and Sakellariadou, Mairi},
  collaboration = {LISA Cosmology Working Group},
  year    = 2019,
  journal = {JCAP},
  volume  = {2019},
  number  = {07},
  pages   = {024},
  doi     = {10.1088/1475-7516/2019/07/024}
}

@article{Ezquiaga:2017dark,
  title   = {Dark Energy after {GW}170817: {D}ead Ends and the Road Ahead},
  author  = {Ezquiaga, Jose Mar{\'i}a and Zumalac{\'a}rregui, Miguel},
  year    = 2017,
  journal = {Phys. Rev. Lett.},
  volume  = {119},
  number  = {25},
  pages   = {251304},
  doi     = {10.1103/PhysRevLett.119.251304}
}

@article{Ferreira:2022forecasting,
  title   = {Forecasting {$F(Q)$} Cosmology with {$\Lambda$CDM} Background
             Using Standard Sirens},
  author  = {Ferreira, Jos{\'e} and Barreiro, Tiago and Mimoso, Jos{\'e}
             and Nunes, Nelson J.},
  year    = 2022,
  journal = {Phys. Rev. D},
  volume  = {105},
  number  = {12},
  pages   = {123531},
  doi     = {10.1103/PhysRevD.105.123531}
}

@article{Foreman-Mackey:2013emcee,
  title   = {emcee: {T}he {MCMC} Hammer},
  author  = {Foreman-Mackey, Daniel and Hogg, David W. and Lang, Dustin
             and Goodman, Jonathan},
  year    = 2013,
  journal = {Publ. Astron. Soc. Pac.},
  volume  = {125},
  pages   = {306},
  doi     = {10.1086/670067}
}

@article{Nishizawa:2018generalized,
  title   = {Generalized Framework for Testing Gravity with
             Gravitational-Wave Propagation. {I}. {F}ormulation},
  author  = {Nishizawa, Atsushi},
  year    = 2018,
  journal = {Phys. Rev. D},
  volume  = {97},
  number  = {10},
  pages   = {104037},
  doi     = {10.1103/PhysRevD.97.104037}
}

@article{Punturo:2010einstein,
  title   = {The {E}instein {T}elescope: A Third-Generation Gravitational
             Wave Observatory},
  author  = {Punturo, M. and others},
  year    = 2010,
  journal = {Class. Quant. Grav.},
  volume  = {27},
  pages   = {194002},
  doi     = {10.1088/0264-9381/27/19/194002}
}

@article{Scolnic:2022pantheon,
  title   = {The Pantheon+ Analysis: {T}he Full Data Set and Light-Curve
             Release},
  author  = {Scolnic, D. and others},
  year    = 2022,
  journal = {Astrophys. J.},
  volume  = {938},
  pages   = {113},
  doi     = {10.3847/1538-4357/ac8b7a}
}

@article{Bluhm2005,
  title   = {Spontaneous {Lorentz} Violation, {Nambu-Goldstone} Modes, and
             Gravity},
  author  = {Bluhm, R. and Kosteleck{\'y}, V. A.},
  year    = 2005,
  journal = {Phys. Rev. D},
  volume  = {71},
  pages   = {065008},
  doi     = {10.1103/PhysRevD.71.065008}
}

@article{Schutz1986,
  title   = {Determining the Hubble Constant from Gravitational Wave
             Observations},
  author  = {Schutz, B. F.},
  year    = 1986,
  journal = {Nature},
  volume  = {323},
  pages   = {310},
  doi     = {10.1038/323310a0}
}

@article{HolzHughes2005,
  title   = {Using Gravitational-Wave Standard Sirens},
  author  = {Holz, Daniel E. and Hughes, Scott A.},
  year    = 2005,
  journal = {Astrophys. J.},
  volume  = {629},
  pages   = {15},
  doi     = {10.1086/431341}
}

@article{Xu:2026bumblebee,
  title    = {Bumblebee Cosmology: {T}he {FLRW} Solution and the {CMB}
              Temperature Anisotropy},
  author   = {Xu, Rui and Xu, Dandan and Andersson, Lars and Amaro Seoane,
              Pau and Shao, Lijing},
  year     = 2026,
  journal  = {Front. Phys.},
  volume   = {21},
  number   = {3},
  pages    = {036201},
  doi      = {10.15302/frontphys.2026.036201}
}

@article{Bertolami:2005vacuum,
  title   = {Vacuum Solutions of a Gravity Model with Vector-Induced
             Spontaneous {Lorentz} Symmetry Breaking},
  author  = {Bertolami, O. and P{\'a}ramos, J.},
  year    = 2005,
  journal = {Phys. Rev. D},
  volume  = {72},
  number  = {4},
  pages   = {044001},
  doi     = {10.1103/PhysRevD.72.044001}
}

@article{Capelo:2015cosmologicala,
  title   = {Cosmological Implications of Bumblebee Vector Models},
  author  = {Capelo, Diogo and P{\'a}ramos, Jorge},
  year    = 2015,
  journal = {Phys. Rev. D},
  volume  = {91},
  number  = {10},
  pages   = {104007},
  doi     = {10.1103/PhysRevD.91.104007}
}

@article{Gonzalez-Espinoza:2026cosmological,
  title   = {Cosmological Implications of Bumblebee Theory on an {FLRW}
             Background},
  author  = {Gonzalez-Espinoza, Manuel and Panotopoulos, Grigorios and
             Tello-Ortiz, Francisco},
  year    = 2026,
  journal = {Fortschr. Phys.},
  volume  = {74},
  number  = {4},
  pages   = {e70110},
  doi     = {10.1002/prop.70110}
}

@article{Guiomar:2014astrophysical,
  title   = {Astrophysical Constraints on the Bumblebee Model},
  author  = {Guiomar, Gon{\c c}alo and P{\'a}ramos, Jorge},
  year    = 2014,
  journal = {Phys. Rev. D},
  volume  = {90},
  number  = {8},
  pages   = {082002},
  doi     = {10.1103/PhysRevD.90.082002}
}

@article{Jesus:2019riccia,
  title   = {Ricci Dark Energy in Bumblebee Gravity Model},
  author  = {Jesus, W. D. R. and Santos, A. F.},
  year    = 2019,
  journal = {Mod. Phys. Lett. A},
  volume  = {34},
  number  = {22},
  pages   = {1950171},
  doi     = {10.1142/S0217732319501712}
}

@article{Khodadi:2023constraining,
  title   = {Constraining the {Lorentz}-Violating Bumblebee Vector Field
             with Big Bang Nucleosynthesis and Gravitational
             Baryogenesis},
  author  = {Khodadi, Mohsen and Lambiase, Gaetano and Sheykhi, Ahmad},
  year    = 2023,
  journal = {Eur. Phys. J. C},
  volume  = {83},
  number  = {5},
  pages   = {386},
  doi     = {10.1140/epjc/s10052-023-11546-3}
}

@article{Khodadi:2025primordial,
  title   = {Primordial Gravitational Waves from Spontaneous {Lorentz}
             Symmetry Breaking},
  author  = {Khodadi, Mohsen and Lambiase, Gaetano and Mastrototaro,
             Leonardo and Poddar, Tanmay Kumar},
  year    = 2025,
  journal = {Phys. Lett. B},
  volume  = {867},
  pages   = {139597},
  doi     = {10.1016/j.physletb.2025.139597}
}

@article{Lai:2025stability,
  title   = {Stability Analysis of Cosmological Perturbations in the
             Bumblebee Model: {P}arameter Constraints and Gravitational
             Waves},
  author  = {Lai, Xiao-Bin and Dong, Yu-Qi and Fan, Yu-Zhi and Liu,
             Yu-Xiao},
  year    = 2026,
  journal = {Phys. Rev. D},
  volume  = {113},
  number  = {4},
  pages   = {044003},
  eprint  = {2509.13958},
  archivePrefix = {arXiv},
  doi     = {10.1103/PhysRevD.113.044003}
}

@article{Liang:2022polarizations,
  title   = {Polarizations of Gravitational Waves in the Bumblebee
             Gravity Model},
  author  = {Liang, Dicong and Xu, Rui and Lu, Xuchen and Shao, Lijing},
  year    = 2022,
  journal = {Phys. Rev. D},
  volume  = {106},
  number  = {12},
  pages   = {124019},
  doi     = {10.1103/PhysRevD.106.124019}
}

@article{Maluf:2021bumblebee,
  title   = {Bumblebee Field as a Source of Cosmological Anisotropies},
  author  = {Maluf, R. V. and Neves, Juliano C. S.},
  year    = 2021,
  journal = {JCAP},
  volume  = {2021},
  number  = {10},
  pages   = {038},
  doi     = {10.1088/1475-7516/2021/10/038}
}

@article{Neves:2023kasner,
  title   = {Kasner Cosmology in Bumblebee Gravity},
  author  = {Neves, Juliano C. S.},
  year    = 2023,
  journal = {Ann. Phys.},
  volume  = {454},
  pages   = {169338},
  doi     = {10.1016/j.aop.2023.169338}
}

@misc{Nilsson:2025bumblebee,
  title  = {Bumblebee Gravity -- {L}essons from Perturbation Theory},
  author = {Nilsson, Nils A.},
  year   = 2025,
  eprint = {2510.13135},
  archivePrefix = {arXiv},
  note   = {Proceedings of the Tenth Meeting on CPT and Lorentz Symmetry
            (CPT'25)}
}

@article{Sarmah:2025anisotropic,
  title   = {Anisotropic Cosmology in Bumblebee Gravity Theory},
  author  = {Sarmah, Pranjal and Goswami, Umananda Dev},
  year    = 2025,
  journal = {Phys. Dark Univ.},
  volume  = {49},
  pages   = {102057},
  doi     = {10.1016/j.dark.2025.102057}
}

@misc{Siquieri:2026cosmic,
  title  = {Cosmic Evolution from {Lorentz}-Violating Bumblebee Dynamics
            and {Tsallis} Holographic Dark Energy},
  author = {Siquieri, E. M. and Cabral, D. S. and Santos, A. F.},
  year   = 2026,
  eprint = {2602.02094},
  archivePrefix = {arXiv},
  note   = {Accepted for publication in Eur.\ Phys.\ J.\ C}
}

@article{VanDeBruck:2026nogo,
  title   = {A No-Go Theorem in Bumblebee Vector-Tensor Cosmology},
  author  = {Van de Bruck, Carsten and Gorji, Mohammad Ali and Nilsson,
             Nils A. and Pookkillath, Masroor C. and Yamaguchi, Masahide},
  year    = 2026,
  journal = {JCAP},
  volume  = {2026},
  number  = {07},
  pages   = {043},
  doi     = {10.1088/1475-7516/2026/07/043}
}

@article{Zhu:2024bumblebee,
  title  = {Bumblebee Cosmology: {T}ests Using Distance- and
            Time-Redshift Probes},
  author = {Zhu, Xincheng and Xu, Rui and Xu, Dandan},
  year   = 2025,
  journal = {Phys. Dark Univ.},
  volume  = {50},
  pages   = {102127},
  eprint = {2411.18559},
  archivePrefix = {arXiv},
  doi    = {10.1016/j.dark.2025.102127}
}

@article{Kostelecky:1989spontaneous,
  title   = {Spontaneous Breaking of {Lorentz} Symmetry in String Theory},
  author  = {Kosteleck{\'y}, V. Alan and Samuel, Stuart},
  year    = 1989,
  journal = {Phys. Rev. D},
  volume  = {39},
  pages   = {683},
  doi     = {10.1103/PhysRevD.39.683}
}

@article{Kostelecky:2004gravity,
  title   = {Gravity, {Lorentz} Violation, and the Standard Model},
  author  = {Kosteleck{\'y}, V. Alan},
  year    = 2004,
  journal = {Phys. Rev. D},
  volume  = {69},
  number  = {10},
  pages   = {105009},
  doi     = {10.1103/PhysRevD.69.105009}
}

@article{Kostelecky:2009gravity,
  title   = {Gravity from Spontaneous {Lorentz} Violation},
  author  = {Kosteleck{\'y}, V. Alan and Potting, Robertus},
  year    = 2009,
  journal = {Phys. Rev. D},
  volume  = {79},
  number  = {6},
  pages   = {065018},
  doi     = {10.1103/PhysRevD.79.065018}
}

@article{Colladay:1997cpt,
  title   = {{CPT} Violation and the Standard Model},
  author  = {Colladay, Don and Kosteleck{\'y}, V. Alan},
  year    = 1997,
  journal = {Phys. Rev. D},
  volume  = {55},
  number  = {11},
  pages   = {6760--6774},
  doi     = {10.1103/PhysRevD.55.6760}
}

@article{Colladay:1998lorentzviolating,
  title   = {{Lorentz}-Violating Extension of the Standard Model},
  author  = {Colladay, D. and Kosteleck{\'y}, V. Alan},
  year    = 1998,
  journal = {Phys. Rev. D},
  volume  = {58},
  number  = {11},
  pages   = {116002},
  doi     = {10.1103/PhysRevD.58.116002}
}

@article{Bailey:2006signals,
  title   = {Signals for {Lorentz} Violation in Post-{Newtonian} Gravity},
  author  = {Bailey, Quentin G. and Kosteleck{\'y}, V. Alan},
  year    = 2006,
  journal = {Phys. Rev. D},
  volume  = {74},
  number  = {4},
  pages   = {045001},
  doi     = {10.1103/PhysRevD.74.045001}
}

@article{Jimenez:2018coincident,
  title   = {Coincident General Relativity},
  author  = {Jim{\'e}nez, Jose Beltr{\'a}n and Heisenberg, Lavinia and
             Koivisto, Tomi},
  year    = 2018,
  journal = {Phys. Rev. D},
  volume  = {98},
  number  = {4},
  pages   = {044048},
  doi     = {10.1103/PhysRevD.98.044048}
}

@article{Jimenez:2020cosmology,
  title   = {Cosmology in {$f(Q)$} Geometry},
  author  = {Jim{\'e}nez, Jose Beltr{\'a}n and Heisenberg, Lavinia and
             Koivisto, Tomi S. and Pekar, Simon},
  year    = 2020,
  journal = {Phys. Rev. D},
  volume  = {101},
  number  = {10},
  pages   = {103507},
  doi     = {10.1103/PhysRevD.101.103507}
}

@article{Ayuso:2021observational,
  title   = {Observational Constraints on Cosmological Solutions of
             {$f(Q)$} Theories},
  author  = {Ayuso, I{\~n}igo and Lazkoz, Ruth and Salzano, Vincenzo},
  year    = 2021,
  journal = {Phys. Rev. D},
  volume  = {103},
  number  = {6},
  pages   = {063505},
  doi     = {10.1103/PhysRevD.103.063505}
}

@article{Barros:2020testing,
  title   = {Testing {$F(Q)$} Gravity with Redshift Space Distortions},
  author  = {Barros, B. J. and Barreiro, T. and Koivisto, T. and
             Nunes, N. J.},
  year    = 2020,
  journal = {Phys. Dark Univ.},
  volume  = {30},
  pages   = {100616},
  doi     = {10.1016/j.dark.2020.100616}
}

@misc{Ferreira:2022repo,
  title  = {Forecasting {$f(Q)$} Cosmology with Standard Sirens},
  author = {Ferreira, J. P. M. V.},
  year   = 2022,
  howpublished = {\url{https://github.com/jpmvferreira/forecasting-FQ-cosmology-with-SS}},
  note   = {Reference implementation of the mock-catalogue and MCMC
            pipeline adapted for this work}
}
\end{document}